\documentclass[12pt,useAMS,useAASTEX,usenatbib]{aastex}
\usepackage{emulateapj5}
\usepackage{times}

\makeatletter
\newenvironment{inlinetable}{%
\def\@captype{table}%
\noindent\begin{minipage}{0.98\linewidth}\begin{center}\footnotesize}
{\end{center}\end{minipage}}

\newenvironment{inlinefigure}{%
\def\@captype{figure}%
\noindent\begin{minipage}{0.999\linewidth}\begin{center}}
{\end{center}\end{minipage}}
\makeatother

\newcommand{\ca}{near-IR Ca\,{\sc ii} triplet}
\newcommand{\cat}{CaT}
\newcommand{\cats}{CaT$^*$}
\newcommand{\pat}{PaT}
\newcommand{\mgi}{Mg\,{\sc i}}

\newcommand{\msun}{\mbox{M$_\odot$}}

\newcommand{\kms}{km~s$^{-1}$}

\defcitealias{centhesis}{CEN02}
\defcitealias{saglia02}{SAG02}
\defcitealias{vazdekis03}{VAZ03}

\shorttitle{The Ca\,{\sc ii} - $\sigma$ relation in bulges of spiral galaxies}
\shortauthors{Falc\'on-Barroso et al.}

\begin{document}

\title{The near-IR Ca {\sc ii} triplet - $\sigma$ relation for bulges of spiral galaxies}

\author{Jes\'us Falc\'on-Barroso and Reynier F. Peletier\altaffilmark{1}} 
\affil{School of Physics \& Astronomy. University of Nottingham. Nottingham. NG7 2RD. United Kingdom}
\altaffiltext{1}{also CRAL, Observatoire de Lyon, F-69561 St-Genis Laval cedex, France}
\author{Alexandre Vazdekis and Marc Balcells}
\affil{Instituto de Astrof\'\i sica de Canarias, E-38200 La Laguna, Tenerife, Spain}

\begin{abstract}
We present measurements of the near-infrared Calcium\,{\sc ii} triplet (\cat, \cats),  
Paschen (\pat) and Magnesium (\mgi) indices for a well-studied sample of 19 bulges of 
early to intermediate spiral galaxies. We find that both the \cats\ and \cat\ indices decrease
with central velocity dispersion $\sigma$  with small scatter. This dependence is similar to
that recently found by \citet{centhesis} for elliptical galaxies, implying an uniform \cats\ -- $\sigma$ 
relation that applies to galaxies from ellipticals to intermediate-type spirals. The decrease of 
\cat\ and \cats\ with $\sigma$ contrasts with the well-known increase of another $\alpha$-element 
index, Mg$_2$, with $\sigma$. We discuss the role of Ca underabundance ([Ca/Fe]$<$0) and IMF 
variations in the onset of the observed relations.  
\end{abstract}

\keywords{galaxies: bulges --- galaxies: fundamental parameters --- galaxies: abundances}

\section{Introduction} 
\label{Sec:Introduction} 
Scaling relations provide powerful insights on the formation and evolution of early-type
galaxies. Besides linking internal structure and dynamics through the Fundamental Plane 
\citep{dd87}, they allow to link these properties with those of stellar populations. The  most
studied scaling relation between population and dynamics is the Mg$_2$ - $\sigma$ relation
\citep{terlevich81}, which  shows that the Mg$_2$ line index increases with galaxy mass. This
scaling relation may be used to study whether bulge populations are younger than those of
elliptical galaxies, with the basic  assumption  that elliptical galaxies in clusters are old. 
However, stellar population studies in the wavelength region of the Mg$_2$  index
\citep{trager00,kuntschner00} indicate that Mg$_2$ is affected by age and metallicity; that Mg is
overabundant w.r.t. Fe in bright galaxies; and, therefore, that the Mg$_2$--$\sigma$  relation for
ellipticals is not necessarily a relation for old galaxies \citep{trager00}. 

The \ca\ at 8500 \AA\ is a bright atomic feature that can easily be studied with modern 
detectors. It has been used in the past to analyse stellar populations  (e.g.
\citealt{carter86,terlevich90,olszewski91,bica98}). Recently, a series of papers have  been
published describing a new stellar library that is used to make a thorough calibration of the
Ca\,{\sc ii} triplet and other features in the same spectral region in terms of stellar
parameters \citep{cen1,cen2,cen3}. A followup paper (\citealt{vazdekis03}, hereafter VAZ03)
presents the behaviour of these features for single stellar populations as predicted by recent
evolutionary synthesis models. \citet[ hereafter CEN02]{centhesis}, have used this calibration to 
study the \ca\ in a sample of elliptical galaxies.  They find that \cats, the Ca triplet index 
corrected for the presence of Paschen lines, decreases with $\sigma$, and  that, in the most
luminous galaxies, \cats\ is much lower than predicted by models.  The decrease of \cats\ with
$\sigma$ may be a result of a dwarf-enhanced IMF or a low Ca/Fe abundance ratio for giant
ellipticals.  In a similar  study, \citet[ hereafter SAG02]{saglia02} find a similar
underabundance of \cat\ w.r.t.  models in luminous ellipticals, but do not see any dependence of
\cats\ with $\sigma$.  

Here we analyse a sample of galactic bulges to compare the trends of Ca with $\sigma$ in bulges to
those found in ellipticals.  Different abundance ratios might be expected if bulges and
ellipticals have  different formation histories. The study offers an opportunity to see whether
\cats\ -- $\sigma$ relation may be established in general for the spheroidal components of
galaxies. 

\section{Observations \& Data Reduction}
\label{Sec:Obsdata}

Nineteen early- to intermediate-type galaxies (S0 to Sbc), from the sample  defined by
\citet{bp94},  were observed using the ISIS double spectrograph at the  4.2m William Herschel
Telescope of the Observatorio del Roque de los Muchachos  at La Palma in 1997 July 11-13. 
Minor-axis spectra were obtained in the 8360 - 9170~\AA\ range,  at 1.74 \AA\ spectral resolution
(FWHM of  the arc lines). The slit width was 1.2\arcsec\ to match the seeing at the time of the
observations. A detailed summary of the observations and data reduction can be found in
\citet{fpb02}. We measured the recently defined line-strength indices \cats, \cat, \pat\ \citep{cen1} 
and  \mgi\ \citepalias{centhesis}, on the flux calibrated spectra. The main advantages of these new 
definitions over the previous ones (e.g. \citealt{armandroff88,diaz89,rutledge97}) are that they have 
been designed to ease problems in the definition of the continuum bands, the contamination of the 
Paschen lines, and the sensitivity to the S/N ratio and velocity dispersion (see \citealt{cen1} for 
more details). In order to explore the dependence of the relation of line-strength and velocity 
dispersion on aperture, we measured the 4 line-strength indices in three different apertures 
(r$_{\rm eff}$, r$_{\rm eff}$/2, 1.2 x 4 arcsec$^2$). The signal-to-noise ratios for the different 
apertures range from 67 to 256, 55 to 232 and 75 to 244 per angstrom, respectively.  $K$-band 
effective radii were taken from  Table 1 of \citet{fpb02} and converted to the minor axis using 
high-spatial resolution  $H$-band ellipticity profiles from \citet{balcells02}. For a number of our 
bulges it was not possible to accurately measure the index at aperture sizes $<$ r$_{\rm eff}$/2 
due to its size and seeing conditions. The difference at smaller apertures, however, should be minor 
since the \ca\ depends very little with the size of the aperture. For brevity, only the r$_{\rm eff}$, 
and 1.2 x 4 apertures are shown in Fig.~\ref{Fig:indexsigma}. The measured values for the 
r$_{\rm eff}$/2 aperture have been included in the electronic version of Table~\ref{Tab:Data}.  Kinematic 
parameters were derived using FOURFIT developed by \citet{vdm93}. A gaussian fit to the LOSVD was performed 
to obtain mean radial velocities and velocity dispersions. We have made use of a single-age, 
single-metallicity synthetic model ([Fe/H]=0.0, t=6.31 Gyr)  from \citetalias{vazdekis03} as a kinematic 
template. This procedure has proven to be very  effective in reducing template mismatch and produces as 
good results as the best stellar  templates in this wavelength domain \citep{fbpa03}. Tabulated values for 
the \cats\ index and velocity dispersions along the minor axis of the galaxies can be found in the same 
paper. We broadened all the spectra to 300 \kms, slightly higher than the largest velocity dispersion in our 
sample (291 \kms\ for the central aperture of NGC~5838). This correction is small for the \cats\ index, which 
has shown to be fairly constant with velocity dispersion up to 300 \kms\ \citepalias{vazdekis03}. Errors 
in the indices include contributions from radial velocity uncertainties and Poisson errors 
(see \citealt{fbpa03} for more details). The uncertainty from S/N dominates the errors in the indices. We 
have corrected our indices for differences between the spectro-photometric system of the data and the model 
predictions using a set of 10 stars in common with the stellar library on which the models are based. 
The following offsets are found (Models-Us): $\Delta$\cats\ =0.178$\pm$0.032,
$\Delta$\cat\ =0.205 $\pm$0.027, $\Delta$\mgi\ =$-$0.001$\pm$0.022, $\Delta$\pat\
=0.029$\pm$0.007. 

\begin{inlinetable}
\caption{Fitting Parameters\label{Tab:Fits}}
{\tabcolsep=0.07in
\begin{tabular}{lcccccccc}
\tableline
 Index    &   a     &  $\Delta$a  &	b     & $\Delta$b  &  r$^2$  &   r$_s$   & $\alpha$(r$_s$) &   rms \\
   (1)    &   (2)   &	  (3)	  &	(4)   &    (5)     &	(6)  &     (7)   &	 (8)	   &   (9) \\
\tableline
    ~	  &	~   &	 ~	  &	~     &   1.2 x 4  &	~    &     ~	 &      ~	   &    ~  \\ 
CaT$^*$   &   8.72  &	 0.38	  &   -0.80   &   0.17     &   0.51  &   -0.71   &    7.4E-04	   &   0.09\\
CaT	  &   8.89  &	 0.40	  &   -0.70   &   0.19     &   0.41  &   -0.58   &    8.6E-03	   &   0.07\\
MgI	  &   0.00  &	 0.08	  &    0.11   &   0.04     &   0.13  &    0.20   &    4.2E-01	   &   0.03\\
\tableline
    ~	  &	~   &	 ~	  &	~     & r$_{\rm eff}$  &    ~	 &     ~    &	   ~	   &    ~  \\   
CaT$^*$   &   9.44  &	 0.39	  &   -1.14   &   0.18     &   0.52  &   -0.76   &    1.7E-04	   &   0.11\\
CaT	  &   9.67  &	 0.42	  &   -1.06   &   0.20     &   0.63  &   -0.80   &    4.2E-05	   &   0.07\\
MgI	  &   0.04  &	 0.09	  &    0.10   &   0.04     &   0.13  &    0.22   &    3.6E-01	   &   0.02\\
\tableline
\end{tabular}}
\begin{minipage}{0.99\linewidth}
{\bf \sc Note:} Parameters of the error-weighted least-squares linear fits to
the index - log($\sigma$) relations, defined as $I=a+b \cdot$ log($\sigma$) at 300 \kms\
resolution. Columns 2, 3, 4, 5 show the fitted parameters and their errors. Columns 6, 7, 8 
give the linear correlation coefficients (r), the Spearman rank-order correlation 
coefficients (r$_{s}$) and significance levels ($\alpha$(r$_s$)). Column 9 lists 
the rms around the best fit.
\end{minipage}
\end{inlinetable}

\section{The index - log($\sigma$) relations}
\label{Sec:relations}

The relations of the line-strength indices \cats, \cat, \pat, \mgi\ and velocity 
dispersion are shown in Figure~\ref{Fig:indexsigma}. We present each relation for the 
two aperture sizes described in \S~\ref{Sec:Obsdata} (labeled on the top boxes).  
Early-type spirals (T $\leq 0$) are depicted with filled symbols, and later types (T $>$0) 
with open symbols.  Index values for a range of SSP (single-age, single-metallicity
stellar populations) models at the resolution of the data are shown in the fourth-column 
panels.  Linear least-squares fits are plotted for some of the relations, with best-fit 
coefficients given in Table~\ref{Tab:Fits}. The data values are presented in Table~\ref{Tab:Data}.

A strong anti-correlation exists between the \cat\ and \cats\ indices and log($\sigma$) 
in all apertures.  This is reflected in the low value of the significance levels of the 
Spearman rank-order tests (Table~\ref{Tab:Fits}, column 8), which give the probability of 
a null correlation, and the rms values about the best fit for the three indices. The 
anti-correlation of the \cats\ index with $\sigma$ shown here for bulges is similar to 
that recently found by \citetalias{centhesis} for elliptical galaxies (which was, if at all, 
marginally detected also for ellipticals by \citetalias{saglia02}). 

\begin{inlinefigure}
\centerline{\includegraphics[angle=00, width=0.99\linewidth]{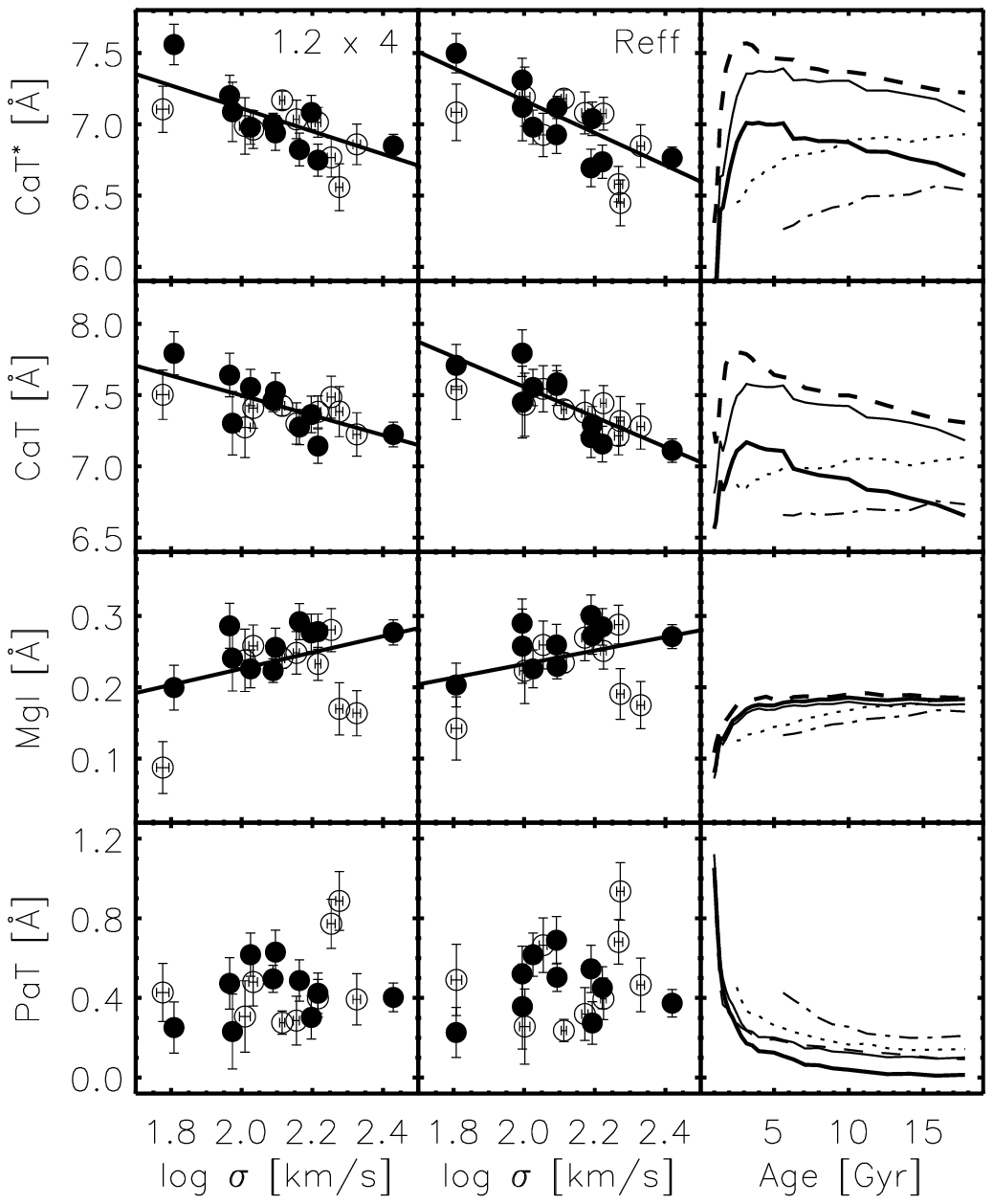}}
\caption{The CaT$^*$, CaT, MgI and PaT index relations with log($\sigma$) for
different apertures. The data has been convolved to a spectral resolution of
300 \kms. Solid circles are early-type bulges (T $\le$ 0), while later-type
bulges (T$>$0) are represented by open circles. Solid lines represent an
error-weighted least-squares fit to the all the points. See Table~\ref{Tab:Fits}
for more details. The last column shows SSP model predictions for different ages and
metallicities, for a Salpeter IMF, for each index. Dashed, solid,
dotted and dash-dotted lines in the last column represent SSP models of
[M/H]=0.2,0.0,-0.4,-0.7 metallicities respectively. The thick solid line, in the last 
column, shows the effect of a higher IMF (i.e. 2.3) for a [M/H]=0.0 model.}
\label{Fig:indexsigma}
\end{inlinefigure}

The Ca\,{\sc ii} triplet  is the first metal line index that is found to anticorrelate with $\sigma$.  
The \mgi\ index, in contrast, shows a positive trend with $\sigma$ as expected from the behaviour of the 
Mg$b$ and Mg$_2$ indices in the optical range (e.g. \citealt{terlevich81}). We have 
investigated whether different apertures would lead to significant differences in 
the index--$\sigma$ relations.  No major difference is found between the gradients of any 
index at each aperture: the slopes and zero-points agree within the errors 
(Table~\ref{Tab:Fits}).  It is interesting to see that none of our bulges give a measurement 
of the \cat\ or \cats\ indices that lies above the predictions of the models 
\citepalias{vazdekis03}.  This result agrees with the model predictions that the \cat\ and 
\cats\ indices tend to saturate for metallicities above [M/H]$\sim$-0.4 \citepalias{vazdekis03},
which are also supported by the data of \citetalias{centhesis} and \citetalias{saglia02}. Given 
that our sample contains bulges from S0 to Sbc, we can study whether the 
index--$\sigma$ relations depend on morphological type (see Fig.~\ref{Fig:indexsigma}). Due 
to our limited number of galaxies, we split the sample in just two groups: one of them
containing galaxies of types S0, S0/a (T$\le$0) and a second group with the Sa to Sbc 
galaxies (T$>$0). No differences are found for any of the relations.

We have compared our index--$\sigma$ relations with those found by \citetalias{centhesis}
measured for an aperture of 2x4 arcsec$^2$. An accurate comparison with \citetalias{saglia02}  is
not possible since their aperture is four times smaller than the smallest we can get
given  the size of some of our bulges.  We perform the comparison with
\citetalias{centhesis} at a  similar aperture size and at the same spectral resolution
(400 \kms). In figure~\ref{Fig:comparison}  we show the comparison of the \cats\ index.
The results are in good agreement at the 1-$\sigma$ level, with the slopes agreeing within
the errors (b[Us]=-0.69$\pm$0.18, b[CEN02]=-0.56$\pm$0.15). The exclusion of one of our
low-velocity dispersion galaxies (NGC~7457), provides an even better agreement in the
slope values (b[Us]=-0.52$\pm$0.19). However the rms of our data around the best fit is
much smaller  (rms[\cats(Us)]=0.14) than the one measured by \citetalias{centhesis}
(rms[\cats(CEN02)=0.28). On the other hand, our sample shows values for the \pat\ index
slightly higher than those of  \citetalias{centhesis} for ellipticals ($\sim$0.03 \AA\ at
400 \kms\ resolution). Previous  studies of the same sample by some of us, studying
deviations of bulges from the FP of ellipticals  and S0s \citep{fpb02} or color-color
diagrams \citep{pb99}, conclude that bulge populations are  old, but could be younger than
ellipticals by up to 2.5 Gyr. The higher \pat\ values, and  correspondingly younger ages
for bulges could also be due to young stars in the centre of our galaxies. However, model
predictions indicate that young populations ($<$1 Gyr), which increase the value of the \pat\
index, also lower the value of \cats, which is opposite to what we find here. Given the small
significance of the offset it is more likely that flux calibration differences contribute to the 
observed shift. For the most extreme case in the sample  (NGC~5965,\pat=0.91 \AA) the high \pat\ 
could be due to star formation induced by the presence of a strong bar in this galaxy 
\citep{kuijken95}. On the other hand, we find no correlation of the \pat\ index with $\sigma$, a 
result supported by \citetalias{centhesis} for Es, and contrary to the strong trend found by  
\citetalias{saglia02}. An explanation for that difference could be in the velocity dispersion 
correction they have applied to each galaxy in their sample. A large correction for more massive 
galaxies could explain the positive correlation of the \pat\ index and the flatness of the \cats\ 
index as a function of velocity dispersion. 

\begin{inlinetable}
\caption{\sc The Data\label{Tab:Data}}
\centering
{\tabcolsep=0.06in
\begin{tabular}{cccccccccccc}
\tableline
 Galaxy &  log($\sigma$)  & \cats &  \cat  &  \mgi  &  \pat  & Aperture \\
   (1)  &      (2)        &  (3)  &   (4)  &   (5)  &	(6)  &   (7)    \\
\tableline
NGC~5326 &  2.224  &  6.897  &  7.238   &  0.252  &  0.367  & r$_{\rm eff}$ \\
         & (0.008) & (0.116) & (0.124)  & (0.026) & (0.105) &               \\    
NGC~5389 &  2.091  &  6.748  &  7.363   &  0.260  &  0.661  & r$_{\rm eff}$ \\
         & (0.010) & (0.132) & (0.141)  & (0.029) & (0.119) &		    \\    
NGC~5422 &  2.221  &  6.559  &  6.951   &  0.286  &  0.422  & r$_{\rm eff}$ \\
         & (0.010) & (0.117) & (0.125)  & (0.026) & (0.106) &		    \\    
NGC~5443 &  2.002  &  7.016  &  7.227   &  0.224  &  0.228  & r$_{\rm eff}$ \\
         & (0.015) & (0.208) & (0.223)  & (0.046) & (0.189) &		    \\    
\hline
\end{tabular}}
\begin{minipage}{0.99\linewidth}
{\bf \sc Note:} The log($\sigma$), \cats, \cat, \mgi\ and \pat\ values for
for our sample of bulges in the different apertures, shown in 
Fig.~\ref{Fig:indexsigma}. Uncertainties in the parameters are shown in
brackets. The following offsets must be added to the tabulated values in
order to obtain the values shown in Fig. 1: $\Delta$\cats=0.178, 
$\Delta$\cat=0.205, $\Delta$\mgi=$-$0.001, $\Delta$\pat=0.029. The 
complete version of this table is in the electronic edition of the Journal.
The printed edition contains only a sample.
\end{minipage}
\end{inlinetable}

\section{Discussion}
\label{Sec:Results}
The origin and abundance evolution of the atomic element Ca has been a matter of debate. Due 
to its nature as an $\alpha$ element, an evolution history similar to that of Mg is predicted 
\citep{ww95}. However several observations reveal a more complex picture where Ca and Mg originate 
and evolve differently. The problem has resuscitated again recently in the light of 
\citetalias{centhesis}, \citetalias{saglia02}. For early-type galaxies, the Ca abundance does 
not follow Mg and there is evidence that it is not enhanced to Fe, from, e.g. measurements of 
the Ca4227 line \citep{vazdekis97,vazdekis01} or the \ca\ \citepalias{centhesis,vazdekis03}. On 
the other hand, the Mg - log($\sigma$) relation has been interpreted as a mass-metallicity 
relation for early-type galaxies \citep{terlevich81}. In the light of the new results obtained 
by \citet{cenletter} and model predictions \citepalias{vazdekis03}, such statement does not necessarily 
hold for the \ca\ - log($\sigma$) relation (see \citealt{worthey98} and \citealt{henry99} for 
extensive reviews).

The large discrepancy between the observed \cats\ and values for a Salpeter-IMF model for galaxies 
with large velocity dispersions can be interpreted as underabundance of Ca w.r.t. the other metals 
in more massive galaxies. Other possibilities are: 1) An IMF biased towards low-mass stars, 2) Composite 
Stellar Populations including a low metallicity component or a young and old population of the same 
metallicity, and 3) Ca depletion. We discuss the advantages/disadvantages of each one of these possible 
explanations in turn.

The problem of the Ca underabundance has been previously addressed by several authors 
(\citealt{oconnell76,mcwill94,vazdekis97,peletier99,vazdekis01}, \citetalias{vazdekis03}). 
It is hard to find an evolutionary scenario that can produce non-solar Ca/Mg values.
Assuming that both Ca and Mg are produced by two "flavours" of SNe in which more massive stars 
(20-40 \msun) produce Mg before ($\approx$2.6 Myr) Ca, generated in less massive stars (12-30 \msun), 
one could explain a small discrepancy in abundances between Mg and Ca \citep{molla00}. However, large 
abundance ratio differences are hard to obtain in this way, and impossible to get for real 
galaxies, for which the formation timescale is probably much larger than a few Myr. 
If we would use $\alpha$-enhanced isochrones (\citealt{salasnich00,vandenberg00},VAZ03), required to 
get the correct [Mg/Fe] ratios in giant ellipticals, assuming a solar-like element partition for the 
[Mg/Ca] and [Fe/Ca], we would predict even larger Ca indices \citepalias{vazdekis03}, increasing 
the discrepancy.

Next we discuss the apparent Ca underabundance at large velocity dispersions in terms of
a  dwarf-enhanced IMF. The universality of the IMF has been a matter of controversy in
the last decades (see \citealt{gilmore01} and \citealt{eisenhauer01} for current reviews
of observational keys in favour and against this statement). IMF time variations among 
ellipticals have been studied by \citet{cenletter} in detail. An advantage of this 
approach is its ability to explain both, the large Mg and the [Mg/Fe] values observed in
giant ellipticals \citep{vazdekis96,cenletter}. The effect of a dwarf-dominated IMF is a
weakening of the \cats\ index  (VAZ03; \citealt{cenletter}). This effect is stronger
at high metallicities.  As also mentioned in \citetalias{saglia02}, there are four
observational drawbacks against this hypothesis: first: indications of an IMF flatter 
than Salpeter in the bulge of our own Galaxy \citep{zoccali00}, second: the
fact that the predicted FeH 9916 \AA\ feature is  significantly stronger than what it is
observed \citep{carter86}, and third: the discrepancy  between visual M/L ratios
(M/L$>$10, \citealt{vazdekis96,maraston98}) and dynamical estimates  M/L$\approx$6
\citep{gerhard01}. An additional drawback is that visual-infrared ($V-K$) colors are just
at the verge of what it is observed in Es, as pointed out by \citet{cenletter}. 

The use of composite stellar populations is analysed by \citetalias{saglia02}. Their
analysis reveals that a combination of populations including a low-metallicity component
was not able to explain the low value of \cats\ and \cat\ together with H$\beta$ for the 
most massive galaxies. Our independent analysis confirms their result.

Ca is known to be highly depleted onto dust in the interstellar medium \citep{crinklaw94}. 
Could it be that it is also depleted in stars? The evidence is not strong. Such depletion in 
stars would require that, during star formation, interstellar dust is recycled less than 
interstellar gas. Moreover, depletion would have to be stronger in more dusty galaxies, so 
one would expect a large scatter in the \cats\ -- $\sigma$ relation due to depletion. At 
least, the scatter for bulges of spirals, which are generally more dusty than ellipticals, 
should be larger. Since this is not seen, we argue that depletion can be excluded as the 
origin of the low \cats\ values.

In summary, many mechanisms can be used to try to explain the apparent underabundance
of Ca for massive galaxies. However none of them alone seems to be able to explain all
the observations. In the light of the discrepancies found between all the possibilities, 
a combination of several effects could be in place. More work is still needed 
to provide an answer to this problem.

\acknowledgements
We thank Javier Cenarro and Javier Gorgas for providing us with software and data 
in electronic form, and for many useful discussions that have helped to improve 
this letter. JFB acknowledges the finantial support from a PPARC studentship. The William 
Herschel Telescope is operated on the island of La Palma by the Isaac Newton Group in the 
Spanish Observatorio del Roque de los Muchachos of the Instituto de Astrof\'\i sica de Canarias.

\begin{inlinefigure}\bigskip
\centerline{\includegraphics[angle=0, width=0.99\linewidth]{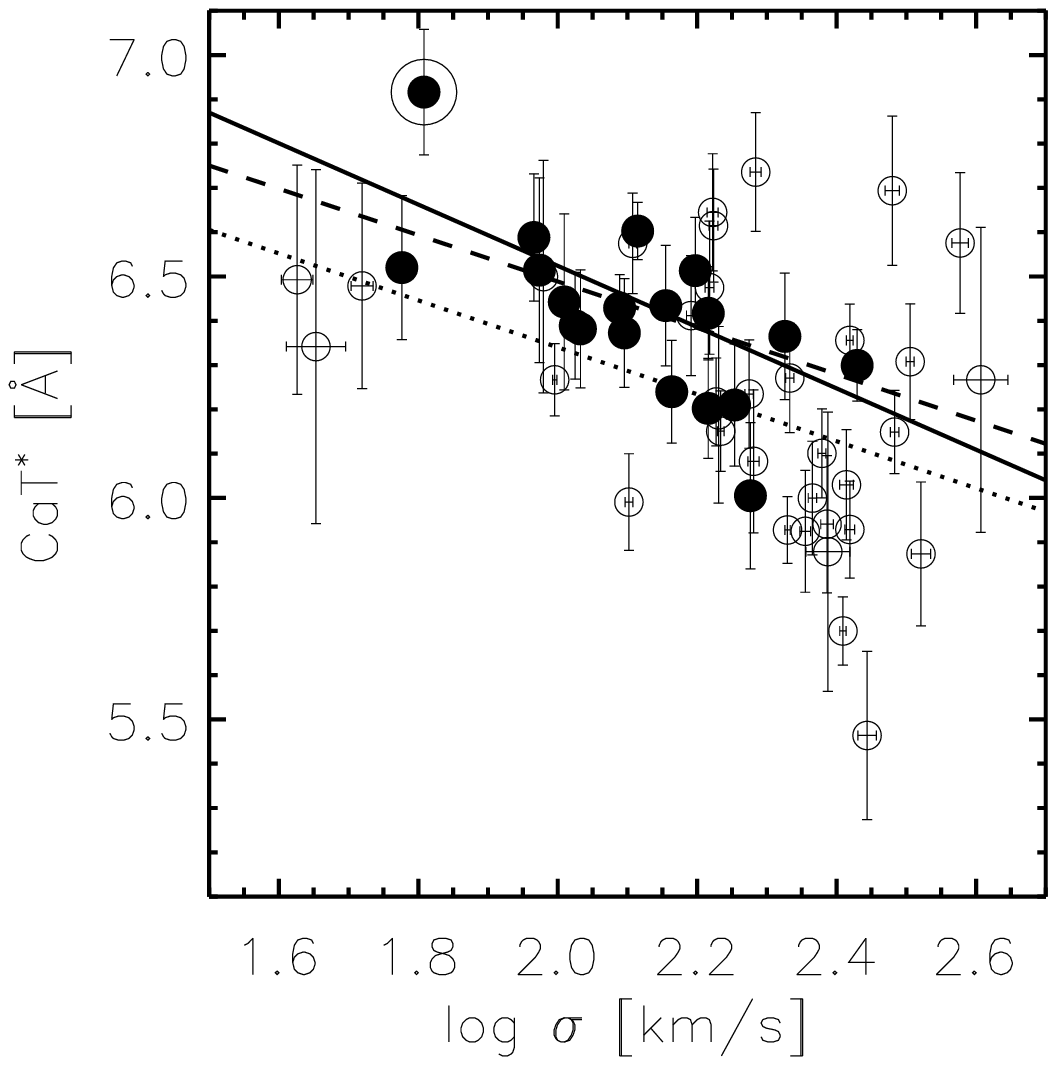}}
\caption[Fig:comparison]{Comparison of the \cats - log($\sigma$) relation
between our sample of bulges and early-type galaxies from \citetalias{centhesis}. Solid
circles and line represent our sample of galaxies and an error-weighted
least-squares linear fit to it. Open circles and dotted line represent the
\citetalias{centhesis} sample of ellipticals and its fit. The dashed line is an error-weighted
least-squares linear fit to our sample excluding the galaxy marked with 
a circle ($\sim$1.8 in log($\sigma$)). In order to compare our results we 
have degraded the resolution of our spectra to 400 \kms.}
\label{Fig:comparison}
\end{inlinefigure}


\end{document}